\documentclass[reprint, aps, prl]{revtex4-1}

\usepackage{amsmath}
\usepackage{amssymb,mathtools}
\usepackage{graphicx}
\usepackage{dcolumn}
\usepackage{bm}
\usepackage{float}
\usepackage{hyperref}
\usepackage{cases}
\usepackage{cleveref}
\usepackage{wasysym}
\usepackage[makeroom]{cancel}
\usepackage{scalerel}
\usepackage{lipsum}
\usepackage{microtype}
\usepackage{comment}

\usepackage{xcolor}
\hypersetup{
	colorlinks,
	linkcolor={red!100!black},
	citecolor={blue!100!black},
	urlcolor={blue!80!black}
}

\DeclareMathOperator{\arctantwo}{arctan2}

\begin{document}

\sloppy
\allowdisplaybreaks
\title{Friction modulation in limbless, three-dimensional gaits and heterogeneous terrains}
\author{Xiaotian Zhang}
\author{Noel Naughton}
\author{Tejaswin Parthasarathy}
\author{Mattia Gazzola}
\email{mgazzola@illinois.edu}
\affiliation{Mechanical Science and Engineering, University of Illinois at Urbana-Champaign, Illinois 61801, USA}

\maketitle

\textbf{Motivated by a possible convergence of terrestrial limbless locomotion strategies ultimately determined by interfacial effects, we show how both 3D gait alterations and locomotory adaptations to heterogeneous terrains can be understood through the lens of local friction modulation. Via an `everything-is-friction' modeling approach, compounded by 3D simulations, the emergence and disappearance of a range of locomotory behaviors observed in nature is systematically explained in relation to inhabited environments. Our approach also simplifies the treatment of terrain heterogeneity, whereby even solid obstacles may be seen as high friction regions, which we confirm against experiments of snakes ‘diffracting’ while traversing rows of posts \cite{Schiebel:2019}, similar to optical waves. We further this optic analogy by illustrating snake refraction, reflection and lens focusing. We use these insights to engineer surface friction patterns and demonstrate passive snake navigation in complex topographies. Overall, our study outlines a unified view that connects active and passive 3D mechanics with heterogeneous interfacial effects to explain a broad set of biological observations, and potentially inspire engineering design.}\\

Limbless locomotion is exhibited by a wide taxonomic range of slender creatures and has been observed in water \cite{Gazzola:2014a}, land \cite{Gray:1946, Gans:1962, Wiens:2006, Astley:2020b}, and even air \cite{Socha:2002}. While broad principles of aquatic limbless locomotion have been unveiled \cite{Taylor:2003, Liao:2003, Gazzola:2012, Gazzola:2014a, Gazzola:2015, Floryan:2017}, the terrestrial variety remains largely elusive. In snakes, locomotion has been classically modeled via planar gaits on uniform substrates, with body undulations rectified into forward motion via anisotropic friction \cite{Alben:2013, Hu:2009,Hu:2012, Marvi:2012, Cicconofri:2015, Gazzola:2018, Guo:2008, Mahadevan:2004, Hazel:1999, Hu:2009}. However, terrestrial creatures (unlike aquatic ones) can actively negotiate the extent of contact with the environment, by lifting selected body regions. This manifests in  a variety of non-planar, transient, and spatially inhomogeneous gaits whose locomotory outputs, in turn, emerge from the interplay with dirt, sand, mud, rocks or leaves \cite{Jayne:2020, Mosauer:1932, Gray:1946, Jayne:1986, Jayne:1988, Gans:1974}, typical of environments that are non-uniform and themselves poorly physically understood.

\begin{figure*}[!t]
\centering
\includegraphics[width=\textwidth]{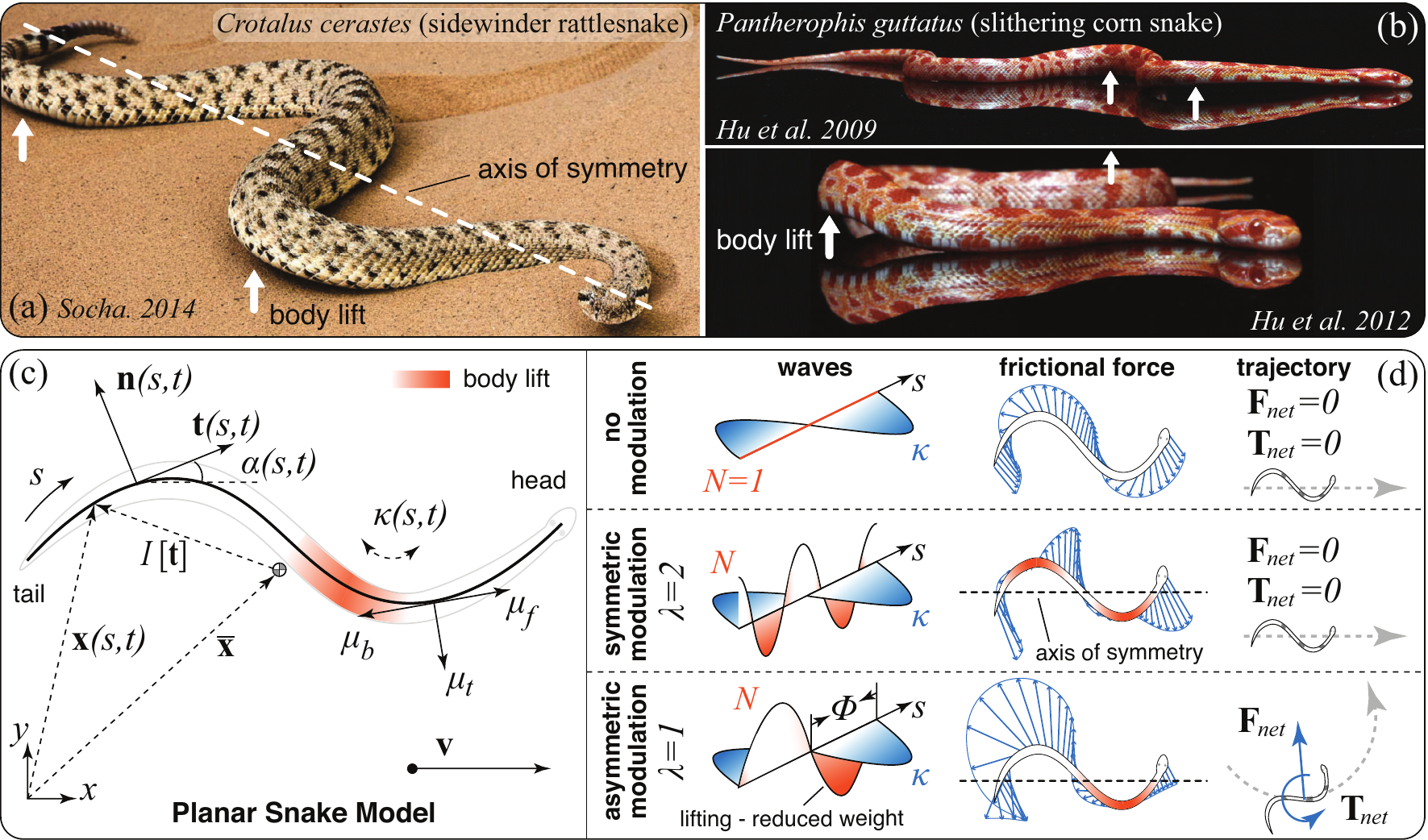}
\vspace{-20pt}
\caption{\footnotesize{Examples of biological snakes employing a lifting body wave in addition to lateral undulation. (a) A sidewinding rattlesnake (\textit{Crotalus cerastes}) asymmetrically lifts up only one side of its body \cite{Socha:2014}. (b) A corn snake (\textit{Pantherophis guttatus}) slithering on a flat surface and symmetrically lifting regions of high curvature on both sides of its body \cite{Hu:2009,Hu:2012}. (c) Schematic of the planar snake model. Note that the arc-length $s$ goes from tail to head to retain consistency with \cite{Hu:2009,Hu:2012}. The local position $\mathbf{x}$ is related to the center of mass $\bar{\mathbf{x}}$ through zero-mean integration function $I[ \mathbf{t}]$ (Methods). (d) Three different stereotypes of body lift: \textit{Top}: Zero body lift leads to classical undulatory planar gaits. \textit{Middle}: Symmetric body lifting, the snake symmetrically lifts both sides of its body \cite{Hu:2009,Jayne:2020}. \textit{Bottom}: Asymmetric body lifting, the snake lifts one side of its body off the ground and maintains the other in contact with the ground. Asymmetric lifting has been well-documented in sidewinding snakes \cite{Mosauer:1932,Gray:1946,Jayne:2020,Astley:2015,Marvi:2014}. Net forces and torques acting on the snake over one undulation period are computed via $\mathbf{F}_{net}=\int^1_0 \int^1_0 \mathbf{F}(s,t) \ ds \ dt$ and $\mathbf{T}_{net}=\int^1_0 \int^1_0 (\mathbf{x}-\bar{\mathbf{x}}) \times \mathbf{F}(s,t) \ ds \ dt$, respectively.}}
\label{Fig:Figure_Model}
\vspace{-15pt}
\end{figure*}

Despite these challenges, a recent convergence between zoologists, physicists, mathematicians, and roboticists has provided  new impetus towards understanding the organization of out-of-plane behaviors \cite{Astley:2015, Marvi:2014, Hu:2009, Jayne:2020, Charles:S-start, Rieser:2021}. Among these, particular attention has been devoted to sidewinding (Fig.~\ref{Fig:Figure_Model}a), whereby snakes can travel at an angle to overall body pose and reorient with neither loss of performance nor kinematic precursors -- features that render sidewinders economical, elusive and versatile dwellers \cite{Secor:1994}. Long puzzling scientists, sidewinding has been recently recapitulated in robot replicas by means of simple actuation templates made of two orthogonal body waves \cite{Astley:2015}, demonstrating steering abilities and ascending of sandy slopes  \cite{Marvi:2014}.

Although insightful, experimental approaches are specialized to given animal/robotic models and there is still a noticeable lack of a broader theoretical perspective able to relate the interplay between gait (body deformation) and frictional environment to locomotory output emergence. Here, motivated by a possible evolutionary convergence of limbless movements ultimately determined by interfacial effects, the roles of both 3D body deformations and environmental heterogeneities are connected through and modeled as planar friction modulations. We then combine theory and simulations to establish an `everything-is-friction' perspective that coherently explains a broad set of observations.

To gain quantitative insight, we generalize a model of forward slithering, first proposed by Hu and Shelley \cite{Hu:2009,Hu:2012}, to encompass a richer variety of behaviors. The model instantiates a snake as a 2D planar curve of lateral curvature $\kappa(s,t)=\epsilon \cos(2 \pi k(s+t))$ with arc-length $s\in[0,1]$, time $t$, amplitude $\epsilon$, and wavenumber $k$, from which midline positions $\mathbf{x}(s,t)$ and orientations $\alpha(s,t)$ follow (Fig.~\ref{Fig:Figure_Model}c, Methods). 
For all quantities, space is scaled on snake's length $L$ and time on wave propagation period $\tau$. Net propulsion forces $\mathbf{F}_{net}$ and torques $\mathbf{T}_{net}$, obtained by integrating friction forces over body length and period, propel the snake. Anisotropic friction forces are described via the Coulomb model $\mathbf{{F}}(s,t) = -N(s,t) \bm{\mu}$, where $\bm{\mu}$ is a function of forward $\mu_f$, transverse $\mu_t$, and backward $\mu_b$ friction coefficients \cite{Hu:2009} (Fig.~\ref{Fig:Figure_Model}, Methods). Of these, $\mu_b$ has little effect \cite{Hu:2012,Alben:2013}, leaving $\mu_t/\mu_f$ as the key characteristic parameter. Thus, system dynamics are governed by the ratio of inertia to friction forces, via the Froude number $Fr = (L/\tau^2)/(g \mu_f)$, with $g$ being gravitational acceleration. In biological and robotic snakes, friction typically dominates with $Fr \leq 1$, and we set $Fr=0.1$ throughout, without lack of generality (SI). Finally, $N(s,t)= \eta \hat{N}(s,t)$ represents local friction modulations due to body lift and weight redistribution ($\eta$ normalization factor, Methods)

The term $\hat{N}(s,t)$ is critical, and much of the model's explanatory power depends on it. Hu et al. \cite{Hu:2009,Hu:2012} set $\hat{N}\sim e^{-\kappa}$ to capture lifting effects at regions of high body curvature in forward slithering snakes (Fig.~\ref{Fig:Figure_Model}b), demonstrating drag reduction and speed increase. Nonetheless, this choice does not capitalize on the opportunity of temporally decoupling lateral and lifting activations, to break symmetry and allow the investigation of locomotory outputs other than forward slithering. While a variety of functions $\hat{N}$ can achieve that (SI), a phase shift in a cosine form consistent with lateral curvature is perhaps the simplest and most natural option. Thus, here we set $\hat{N}(s,t)=\max\{0,A \cos(2 \pi k_l (s+t+\Phi))+1\}$ with $A$ lifting amplitude, $\Phi$ phase offset with lateral wave $\kappa$, and $\lambda = k_l/k$ ratio of lateral to lifting wave numbers. The max function avoids artificial negative weight redistributions.

This parameterization allows us to model and compare stereotypical lifting patterns encountered in nature (Fig.~\ref{Fig:Figure_Model}d). For $A=0$, classic planar undulatory gaits are recovered \cite{Jayne:1986,Hu:2009}. For $|A|>0$ and $\lambda=2$, the snake symmetrically lifts both sides of its body, as in \cite{Hu:2009, Jayne:2020}. In both cases, $\mathbf{F}_{net}=\mathbf{T}_{net}=0$ due to symmetry and the snake can only move forward. If instead $\lambda=1$, the snake lifts only on one side, as seen in sidewinders \cite{Jayne:1986,Marvi:2014}. This breaks friction forces symmetry, allowing maneuvering ($\mathbf{F}_{net}$, $\mathbf{T}_{net}\neq 0$) without changes in the lateral gait $\kappa$.

\begin{figure*}[!th]
\centering
\includegraphics[width=0.99\textwidth]{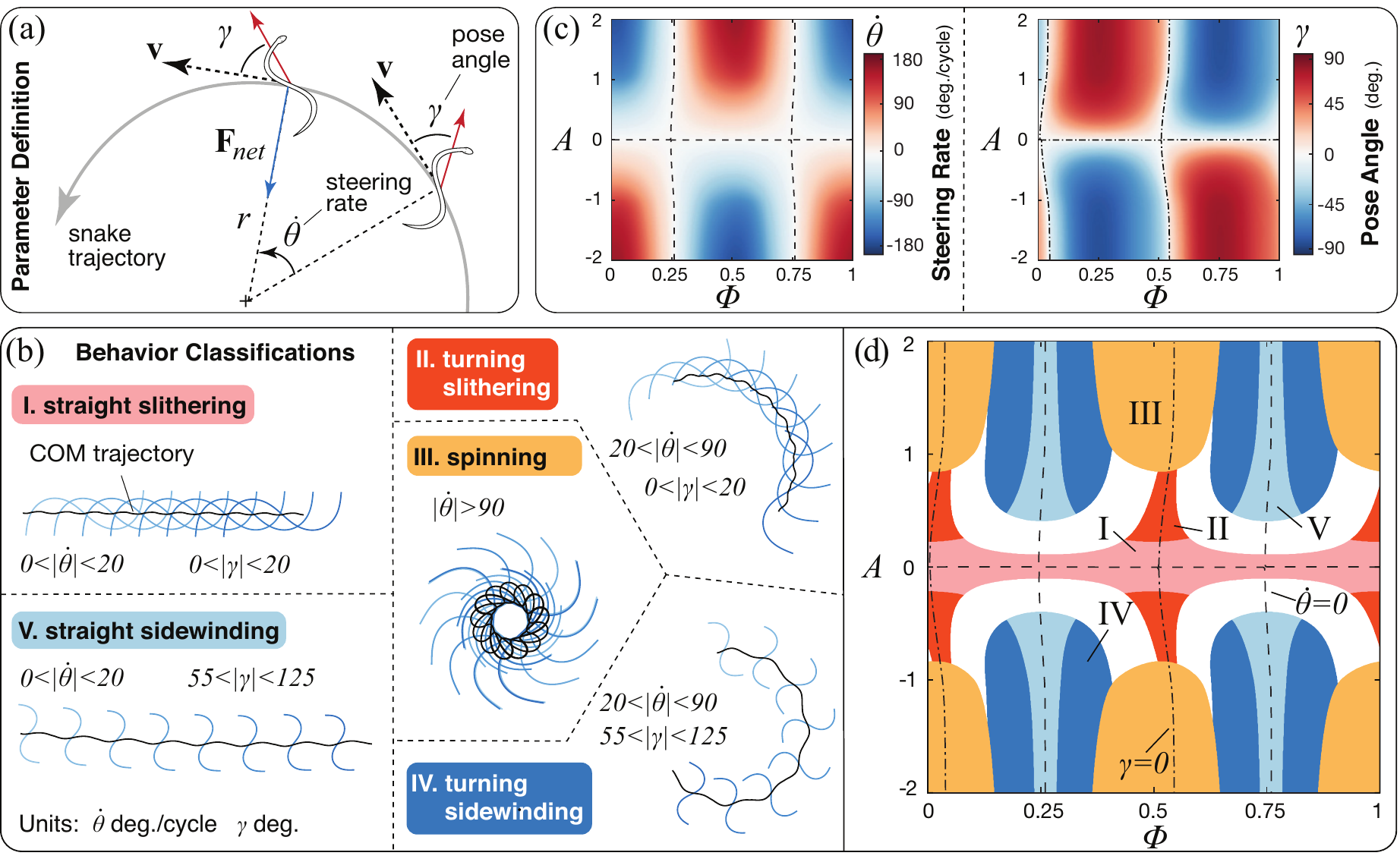}
\caption{\footnotesize{(a) Quantities used to analyze locomotion behavior. Steering rate $\dot{\theta}$ is the time-averaged angular velocity of the snake's center of mass. Pose angle $\gamma$ is the angle between the snake's orientation and velocity direction (Methods). (b) Classifications of qualitatively different locomotion behaviors given $\dot{\theta}$ and $\gamma$ (Movie S1), and based on experimentally observed pose angles of sidewinding snakes \cite{Tingle2020}. Black lines are the snake's center of mass trajectories. (c) Field map of steering rate $\dot{\theta}$ and pose angle $\gamma$ for varying lifting wave amplitude $A$ and phase offset $\Phi$ (for $\mu_t/\mu_f=2$). (d) Phase space of locomotion behaviors available to a snake for the friction ratio $\mu_t/\mu_f=2$. White spaces are transition regimes between different behaviors. Separatrices are zero contours for steering rate (dash line) and pose angle (dash-dot line).}}
\label{Fig:Figure_Phase}
\vspace{-15pt}
\end{figure*}

To investigate the potential of $\lambda=1$ lifting waves for locomotion, we identify the behaviors available to a snake in relation to its frictional environment. We consider first the ratio $\mu_t/\mu_f=2$, which captures the frictional interaction between anisotropic scales and firm uniform substrates, determined for anesthetized snakes \cite{Hu:2009}. Since snakes actively control their scales for grip \cite{Marvi:2012}, $\mu_t/\mu_f=2$ may be considered a lower bound estimate.

We numerically span the $A$--$\Phi$ plane and characterize locomotory outputs by steering rate $\dot{\theta}$ and body pose $\gamma$ (Fig.~\ref{Fig:Figure_Phase}a,b), based on experimentally observed  behaviors \cite{Mosauer:1932, Gray:1946, Jayne:2020, Marvi:2014, Astley:2015}. Key organizing separatrices emerge (Fig.~\ref{Fig:Figure_Phase}c). Along $A=0$ or $\Phi\sim1/4$ and 3/4, the snake can only travel in rectilinear trajectories ($\dot{\theta}=0$), whether it is slithering or sidewinding. At the same time, along $A=0$ or $\Phi\sim0$ and 1/2, the snake is always tangent to its trajectory ($\gamma=0$), whether traveling rectilinearly or turning. Around this underlying structure, locomotion behaviors naturally organize as phases (Fig.~\ref{Fig:Figure_Phase}d). Straight slithering is encountered throughout $\Phi$ for small $A$, with limited turning abilities observed in small regions at $\Phi\sim0$ and 1/2. Sidewinding clusters around $\Phi\sim1/4$ and 3/4 for larger lifting. This explains observations of $\Phi\sim1/4$ in biological sidewinders \cite{Jayne:1986,Mosauer:1932} and empirical robotic demonstrations \cite{Marvi:2014, Astley:2015}: indeed only in the neighborhood of this particular offset (or equivalently $\Phi\sim3/4$) can both linear trajectories and large pose angles co-exist. Finally, spinning in-place \cite{Astley:2015} fills gaps at high liftings.

\begin{figure*}[t]
\centering
\includegraphics[width=0.99\textwidth]{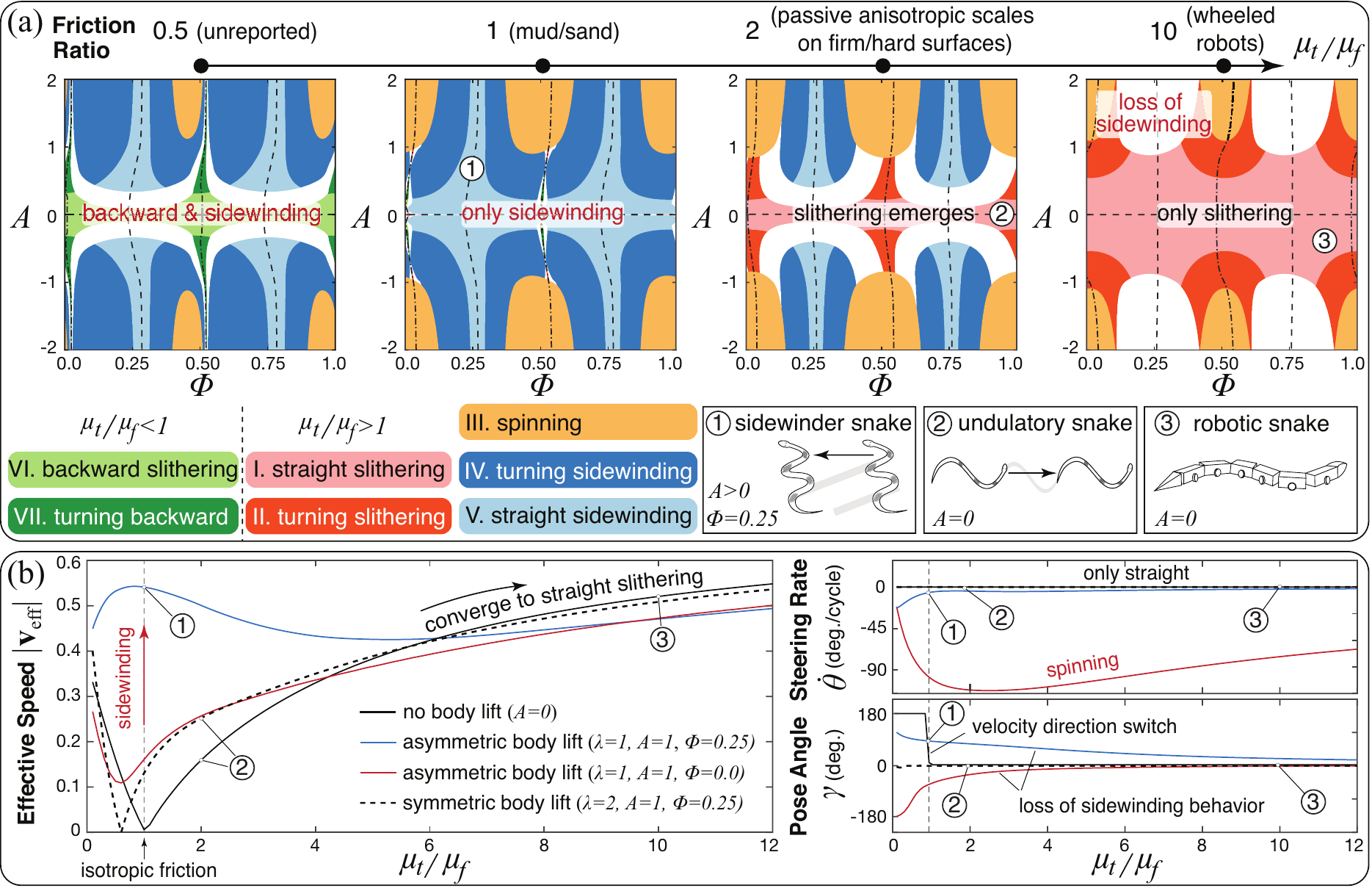}
\caption{\footnotesize{(a) Phase space maps for different anisotropic friction ratios $\mu_t/\mu_f$ (full exploration in SI). Numbered labels indicate the location in terms of friction ratio, lifting amplitude, and phase offset of typical locomotion behaviors: (1) a snake sidewinding in a sandy desert or a tidal mudflat \cite{Jayne:1986}, (2) an undulating snake with no body lift based on measurements from \cite{Hu:2009}, and (3) a wheeled robotic snake where the wheels can be viewed as inducing strong friction anisotropy. Note that for $\mu_t/\mu_f=1$, straight sidewinding gait does not occur for $A=0$ as the snake is unable to produce directional motion when there is no body lift. (b) Effective speed, steering rate, and pose angle of four different lifting behaviors over a range of friction ratios. Different body lifting behaviors lead to large differences in all three quantities at low friction anisotropy ratios while there is a general convergence of behaviors to traveling forward in a straight trajectory as friction anisotropy increases. For isotropic friction, the no lifting case of $A=0$ corresponding to planar slithering has $|\mathbf{v}_\text{eff}|=0$, while asymmetric lifting with $A=1$ and $\Phi=0.25$, corresponding to a sidewinding gait \cite{Astley:2015,Marvi:2014}, exhibits high $|\mathbf{v}_\text{eff}|$.}}
\label{Fig:Figure_Friction}
\vspace{-15pt}
\end{figure*}

To further contextualize these findings, it is useful to investigate how changes in the snake-environment interaction, captured by $\mu_t/\mu_f$, affect phase space organization. As we vary $0.5<\mu_t/\mu_f<10$ in Fig. \ref{Fig:Figure_Friction}a, separatrices are approximately retained, while behavioral outputs drastically remodel, appear and disappear. For example, for $\mu_t/\mu_f< 1$ (a condition not commonly encountered in nature, included here for completeness) slithering is replaced by a new, backward counterpart wherein snakes completely reverse their travel direction. 

\begin{figure*}[!t]
\centering
\includegraphics[width=0.99\textwidth]{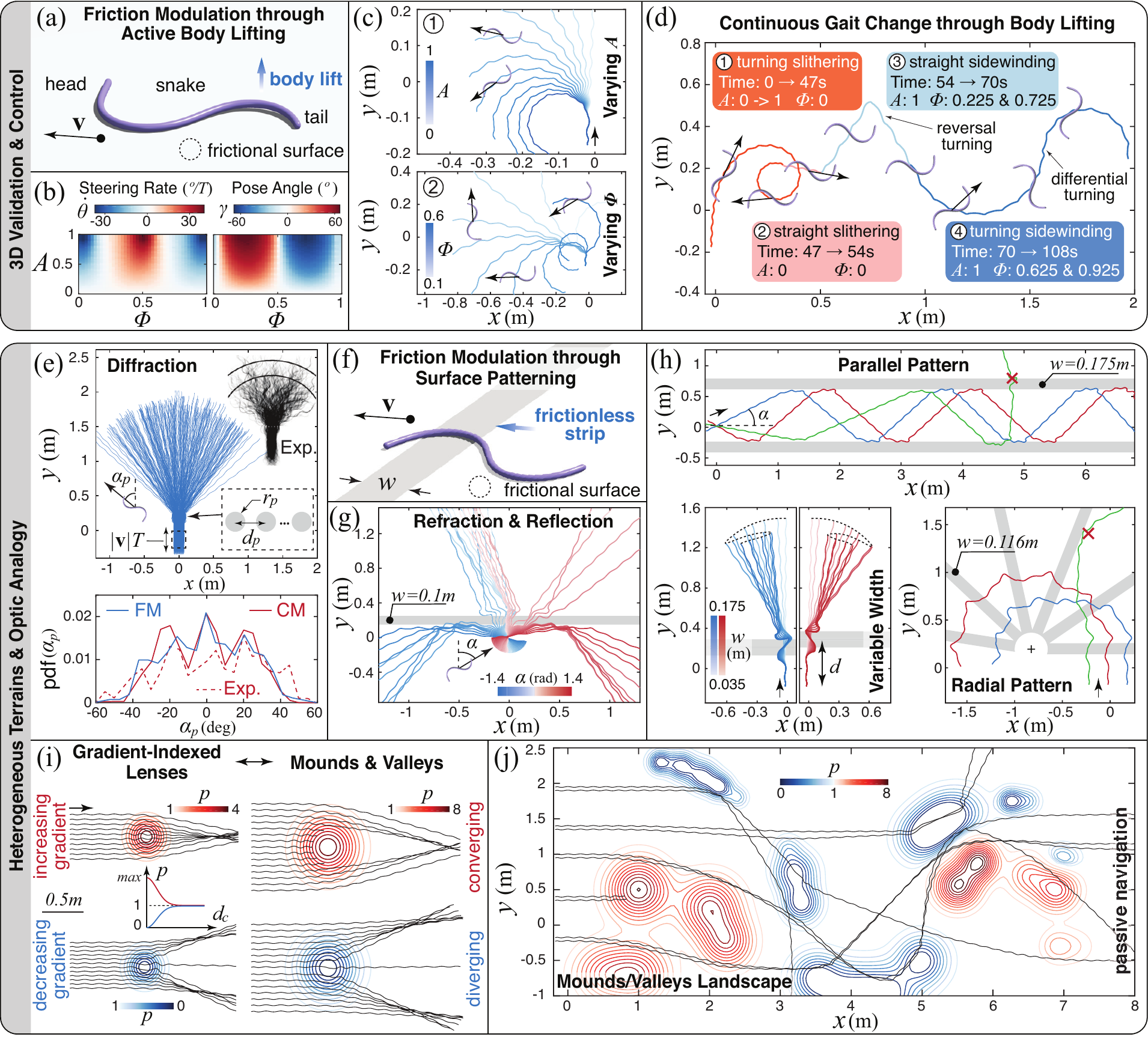}
\caption{ \footnotesize{(a) Schematic of a 3D elastic snake model with internal muscular activation and out-of-plane body lift. (b) Field maps of steering rate and pose angle for $\mu_t / \mu_f=2$. (c) Trajectories for (1) $\Phi = 0.5$ and $A \in [0,1]$, (2) $A=1$ and $\Phi \in [0.1,0.6]$ (full details in SI). (d) Complex trajectories possible by controlling lift amplitude and phase offset. (e) Diffraction pattern and probability density function (pdf) of diffracted angle $\alpha_p$ for simulations of snakes slithering through regularly spaced patches of high friction (friction model --- FM, friction is modulated by scaling the local friction coefficient of the patches by a large factor $p$), and comparison with experimental observations of biological snakes traveling through rigid posts (Exp.) and with collision model (CM) simulations \cite{Schiebel:2019}. (f) A snake moving on a flat surface ($\mu_t / \mu_f=10$) patterned with frictionless strips of width $w$. (g) Snakes encountering a frictionless strip are either reflected or refracted depending on incidence angle $\alpha$. (h) Demonstrations of passive trajectory control through friction surface patterning (additional details for all cases in SI). (i) Snakes interacting with heterogeneous ground features of diameter $d_c$ for both increasing and decreasing friction modulation $p$. (j) Snakes passively meander through an heterogeneous frictional contour map.}}
\label{Fig:Figure_Elastica}
\vspace{-15pt}
\end{figure*}
	
For isotropic friction $\mu_t/\mu_f=1$, planar ($A=0$) and asymmetric lifting ($\lambda=1$) slithering are no longer available, and snakes must instead either sidewind or switch to symmetric lifting ($\lambda=2$) for locomotion. However, sidewinding is found to be significantly faster (Fig~\ref{Fig:Figure_Friction}b), thus proving advantageous in environments such as sandy deserts or mudflats, characterized by low friction ratios on account of their propensity to yield under stress \cite{Maladen:2009,Astley:2020}. This is consistent with sidewinders inhabiting such terrains \cite{Maladen:2009}, while, conversely, slithering snakes are found to adopt sidewinding when encountering sand and mud \cite{Gans:1974,Tingle2020,Jayne:1986}. Further confirming predictions, the application on slithering snakes of cloth `jackets' that eliminate anisotropy severely impairs locomotory performance \cite{Goldman:2010}.

As friction ratios increase ($\mu_t/\mu_f > 2$), we observe a progressive loss of sidewinding behavior (Fig.~\ref{Fig:Figure_Friction}a,b) and a convergence towards slithering, which becomes increasingly faster and eventually, for sufficiently large values ($\mu_t/\mu_f > 10$, e.g. wheeled robots), the only option. This is again consistent, with observations that sidewinding rarely occurs (even in snakes that regularly use it) outside of sandy and muddy terrains \cite{Tingle2020}, and with the fact that sidewinders and slitherers are comparably fast in their respective habitats \cite{Hu:2012}.

Thus, by reducing out-of-plane deformations to waves of active friction modulation, our simple model coherently captures a broad set of experimental observations, providing a mapping between gait, frictional environment and locomotory output. In particular, it corroborates the hypothesis, never mechanistically rationalized, of sidewinding being an adaptation to sandy/muddy contexts \cite{Tingle2020}. Our model mathematically predicts the natural emergence of sidewinding in nearly isotropic environments as a consequence of temporal decoupling between lateral and vertical undulations, and its selection as advantageous in terms of locomotory performance. This perspective recently received notable experimental support with evidence of evolutionary convergence in the ventral skin of sidewinding vipers across world deserts \cite{Rieser:2021}. Their skin indeed evolved from well-documented anisotropic textures in non-sidewinders, to isotropic ones. This, according to our model, maximizes sidewinding locomotion speed (far outperforming other options) and offers a rationale for the observed evolutionary selection (Fig.~\ref{Fig:Figure_Friction}a,b).

Next, we verify our findings in full 3D simulations, whereby a limbless active body is represented as a Cosserat rod (Fig.~\ref{Fig:Figure_Elastica}a) equipped with internal muscular activity and interacting with the substrate through contact and friction \cite{Zhang:2019} (Methods/SI). We then instantiate a snake consistent with \cite{Hu:2009,Hu:2012}, and replace friction modulation waves with torque waves (of the same form), to produce actual body lift (Movie S2).

As can be seen in Fig.~\ref{Fig:Figure_Elastica}b, the 3D model produces steering rates  $\dot{\theta}$ and poses $\gamma$ consistent with Fig. \ref{Fig:Figure_Phase}b, recovering all modes of locomotion in Fig. \ref{Fig:Figure_Phase}d. Examples of slithering and sidewinding ($\lambda=1$) are reported in Fig.~\ref{Fig:Figure_Elastica}c, showing degrees of turning in line with Fig. \ref{Fig:Figure_Phase}d. Further informed by Fig.~\ref{Fig:Figure_Phase}d, a repertoire of linear displacements, wide/tight turns and reversals at varying body poses are then concatenated in Fig.~\ref{Fig:Figure_Elastica}d (Movie S3), illustrating trajectory control in the spirit of \cite{Astley:2015}.

With consistent direct numerical simulations in hand, we switch from a perspective where the snake actively modulates friction to one where locomotory output is instead passively altered by friction patterns on the substrate. The goal is to understand how far our `everything-is-friction' perspective can be pushed to investigate heterogeneous environments comprised of small and large scale 3D features.

We start by considering a recent study proposing an intriguing optical analogy, whereby snake's body undulations and center of mass represent, respectively, `wave' and `particle' nature of light \cite{Rieser:2019,Schiebel:2019}. There, the authors engineer an environment made of seven rigid cylindrical posts aligned with fixed spacing (Fig. \ref{Fig:Figure_Elastica}g), and let both biological \cite{Schiebel:2019} and robotic \cite{Rieser:2019} snakes slither through the posts, propelled by an approximately planar, stereotypical gait. Surprisingly, the snake-posts physical interaction is found to lead to characteristic diffraction patterns. We challenged our approach to reproduce this experiment by taking the drastic step of representing the rigid posts as circular patches of high friction on the ground. As seen in Fig.~\ref{Fig:Figure_Elastica}e, our simulations quantitatively match observed deflection distributions, showing how environmental heterogeneities can be successfully modeled as planar friction patterns,  simplifying treatment.

Motivated by these results, we further explore the connection with optics, to build intuitive understanding of heterogeneous environments, design passive control strategies or anticipate failure modes in robotic applications.

As illustrated in Fig.~\ref{Fig:Figure_Elastica}f-h, a variety of optical effects can be qualitatively reproduced. For example, by patterning a thin low friction strip, we can form an interface, recovering refraction and reflection patterns typical of light transport across two media. Further, we can use this insight and modify the width and spatial arrangement of low fiction strips, to control trajectory deflections, produce U-turns or even guide snakes along a `channel', analogous to optic fiber light transport (Movies S4, S5). 

Similarly, we can imitate light convergence/divergence in gradient-index lenses \cite{Merchand:2012} by simply creating friction gradients as illustrated in Fig.~\ref{Fig:Figure_Elastica}e. This approach informs the modeling of large scale (several body lengths) 3D landscape features, such as mounds and valleys. Indeed, slopes may be seen as unbalancing lateral frictional responses, causing the snake to coast in converging (valleys) or diverging (mounds) patterns. To illustrate this concept and further demonstrate the potential for passive, robust control through friction design, we create a topographic map (Fig. \ref{Fig:Figure_Elastica}f) and challenge snakes initialized at different locations to slither through the map, without altering their gaits. As can be seen, snakes meander through the landscape with about $\sim$50\% of them making it to the other end, with no active control, showing how friction naturally mediates passive adaptivity to deal with heterogeneities in the environment  (Movie S6). 

In summary, through minimal theoretical modeling and 3D simulations, our study contextualizes a broad set of observations, both in the biological and robotic domain, through a unified framework centered around friction modulation, active or passive, uniform or heterogeneous, in 2D or 3D, naturally encountered or engineered. It provides a mathematical argument supporting the convergent evolution of sidewinding gaits, while reinforcing the analogy between limbless terrestrial locomotion and optics, demonstrating its utility for passive trajectory control, with potential applications for bio-inspired engineering.

\vspace{-15pt}
\section*{Methods}
\vspace{-10pt}

\noindent
\textbf{Planar model of friction modulation.}
We adopt the approach of Hu et al. \cite{Hu:2009,Hu:2012} wherein the centerline of a snake of length $L$ is modeled as an inextensible planar curve $\hat{s} \in [0,L]$. The center of mass position and average orientation of the snake are denoted by $\hat{\bar{\mathbf{x}}}(t)$ and $\bar{\alpha}(t)$, respectively, and the local position and orientation of each point along the snake's centerline is computed via $\hat{\mathbf{{x}}}(\hat{s},\hat{t})=\hat{\bar{\mathbf{x}}}(\hat{t})+\textit{I}[\mathbf{t}(\hat{s},\hat{t})]$ and $\alpha(\hat{s},\hat{t})=\bar{\alpha}(\hat{t})+I[\hat{\kappa}(\hat{s},\hat{t})]$, respectively, where $\mathbf{t}(\hat{s},\hat{t})=(\cos\alpha(\hat{s},\hat{t}), \sin\alpha(\hat{s},\hat{t}))$ is the local tangent vector, $\hat{\kappa}(\hat{s},\hat{t})$ is the local curvature and $I[f(\hat{s},\hat{t})]=\int_{0}^{\hat{s}}f(\hat{s}^{\prime},\hat{t})d\hat{s}^{\prime}-\frac{1}{L}\int_{0}^{L}\int_{0}^{\hat{s}}f(\hat{s}^{\prime},\hat{t})d\hat{s}^{\prime}d\hat{s}$ is a mean-zero integration function, which expresses the mathematical machinery that allows us to reconstruct snake's local positions/orientations from center of mass, global orientation and curvature information (Fig. \ref{Fig:Figure_Model}c). Differentiating $\hat{\mathbf{{x}}}(\hat{s},\hat{t})$ twice with respect to time yields
\begin{equation}
	\hat{\mathbf{x}}_{tt}=\hat{\bar{\mathbf{x}}}_{tt}+I[-({\bar{\alpha}}_t+I[\hat{\kappa}_t])^2\mathbf{t}]+I[(\bar{\alpha}_{tt}+I[\hat{\kappa}_{tt}])\mathbf{n}]. 
	\label{Xtt}
\end{equation}
where $\mathbf{n} = (-\sin\alpha, \cos\alpha)$ is the local normal vector. Writing the snake's dynamics as a force balance of internal $\hat{\mathbf{f}}$ and external $\hat{\mathbf{F}}$ forces per unit length yields
\begin{equation}
	\rho \hat{\mathbf{x}}_{tt}(\hat{s},\hat{t})=\hat{\mathbf{F}}(\hat{s},\hat{t})+\hat{\mathbf{f}}(\hat{s},\hat{t}), 
	\label{Newton}
\end{equation}
where $\rho$ is the line density of the snake. We then scale Eqs. \ref{Xtt} and \ref{Newton} by $s={\hat{s}}/ {L}$ and $t={\hat{t}}/ {\tau}$ to non-dimensionalize the system. 

External forces stem entirely from frictional effects captured through the Coulomb friction model, with anisotropy characterized by coefficients in the forward (${\mu_f}$), backward (${\mu_b}$), and transverse (${\mu_t}$) directions. Scaling friction forces such that $\mathbf{F}={\hat{\mathbf{F}}}/ {\rho g \mu_{f}}$ allows us to write the friction force as $\mathbf{{F}}(s,t) = -N(s,t) \bm{\mu}(s,t)$ with 

\noindent
{\small
$$\bm{\mu}(s,t) = \frac{\mu_t}{\mu_f} \left(\mathbf{u}\cdot \mathbf{n}\right)\mathbf{n} + 
					\left[\left(H(\mathbf{u}\cdot \mathbf{t}\right) + \frac{\mu_b}{\mu_f} \left(1 - H(\mathbf{u}\cdot \mathbf{t}\right)\right]
					\left(\mathbf{u}\cdot \mathbf{t} \right) \mathbf{t},$$
\normalsize}%
where $\mathbf{u}(s) = \mathbf{x}_t(s)/\left|\mathbf{x}_t(s)\right|$ is the unit vector associated with the snake's local velocity direction, and $H=1/2(1+\text{sgn}(x))$ is the Heaviside step function used here to distinguish between forward and backwards friction components. Here, $\mu_b/\mu_f=1.5$, in keeping with experimental observations \cite{Hu:2009}. Previous work \cite{Hu:2012} and our preliminary investigations found that static friction effects do not appreciably influence the snake's steady-state behavior, thus we did not consider them here. Moreover, we write the friction modulation wave as $N(s,t)=\eta \hat{N}(s,t)$, where $\eta= 1/ \int^1_0 \hat{N}(s,t) \ ds$ is the normalization constant to conserve the overall weight of the snake. Finally, we set $Fr=0.1$ throughout, consistent with snakes' typically low values and without lack of generality. 

Assuming the total non-dimensionalized internal forces and torques to be zero ($\int^1_0\mathbf{{f}}d{s}=0$ and $\int^1_0(\mathbf{x}-\bar{\mathbf{x}}) \times \mathbf{{f}}d{s}=0$) yields the snake's equation of  motion
{\small
\begin{align}
	& Fr \ \bar{\mathbf{x}}_{tt}(t)=\int^1_0 -N(s,t) \bm{\mu}(s,t) ds
	\label{Linear} \\
	& \begin{multlined}
		Fr \ \bar{\alpha}_{tt}(t) = \frac{1}{J} \int^1_0-(\mathbf{x}-\mathbf{x}) \times N \bm{\mu} \, ds \\ + 
		\frac{Fr}{J} \int^1_0I[\mathbf{n}]\cdot 
		I[\mathbf{t}({\bar{\alpha}}_t+I[\kappa_t])^2] -
		I[\mathbf{t}]\cdot I[\mathbf{t}I[\kappa_{tt}]]ds,
	\end{multlined}
	\label{Angular}
\end{align} 
\normalsize}%
where $J=\int^1_0(\mathbf{x}-\bar{\mathbf{x}})^2ds$ is the moment of inertia. These equations can then be solved for a prescribed non-dimensional curvature $\kappa(s,t)$ and friction scaling term $N(s,t)$.

In all cases considered here, Eqs. \ref{Linear} and \ref{Angular} are numerically solved over 10 undulation periods to allow transient effects from startup to dissipate, and the snake to reach steady state behavior. The snake's locomotion behavior is then analyzed in terms of the pose angle $\gamma$, steering rate $\dot{\theta}$ and effective speed  $|\mathbf{v}_{\text{eff}}|$ which are illustrated in Fig. \ref{Fig:Figure_Phase}a. At steady state, the first trajectory metric that can be computed is the pose angle, which is the angle between the snake's average orientation $\bar{\mathbf{t}}=(\cos \bar{\alpha}, \, \sin \bar{\alpha})$ and its center of mass velocity direction $\bar{\mathbf{u}}$. The average pose angle over one undulation period is defined as $\gamma = \int_{t_0}^{t_1} \arctantwo ((\bar{\mathbf{t}} \times \bar{\mathbf{u}}) \cdot \mathbf{e}_z, \bar{\mathbf{t}} \cdot \bar{\mathbf{u}}) \, dt / \mathcal{T}$ where $\mathcal{T} = \int_{t_0}^{t_1} dt$ and $\mathbf{e}_z$ is the unit vector of out-of-plane axis. Use of the $\arctantwo$ function is required to ensure $\gamma \in (-\pi,\pi]$. Note that $\mathcal{T}$ is for a non-dimensionalized time period, so over one undulation, $\mathcal{T}=1$. Additional trajectory metrics can be computed by considering the snake's center of mass as a particle undergoing planar motion in polar coordinates, $\bar{\mathbf{x}}(t)=(r\cos\theta, r\sin\theta)$, allowing the snake's trajectory to be quantified in terms of its effective velocity $|\mathbf{v}_{\text{eff}}| = \left|\int_{t_0}^{t_1} r \frac{d\theta}{dt} \, \mathbf{u}_\theta \, dt / \mathcal{T} \right|$ and steering rate $\dot{\theta} = \int_{t_0}^{t_1} \frac{d\theta}{dt} \, dt / \mathcal{T}$ (see SI Note 1 for relevant derivations). 

For phase space simulations, a simulation grid was defined with 501 equidistant points in both $A \in [-2,2]$ and $\Phi \in [0,1]$, leading to 251k simulations for each of the friction ratios considered. Simulations were performed on the Bridges supercomputing cluster at the Pittsburgh Supercomputing Center.

\textbf{3D elastic model of snake locomotion.}
3D elastic simulations of snake locomotion were performed in Elastica \cite{Gazzola:2018,Zhang:2019,Naughton:2021} using a Cosserat rod snake model with muscular activation, an approach demonstrated in numerous biophysical applications \cite{Gazzola:2018, Pagan-Diaz:2018, Aydin:2019, Charles:2019b, Zhang:2019, Chang:2020, Wang:2021}. For the Cosserat rod model, we mathematically describe a slender rod by a centerline $\bar{\mathbf{x}}(s, t) \in \mathbb{R}^3$ and a rotation matrix $\mathbf{Q}(s, t)=\{ \bar{\mathbf{d}}_1, \bar{\mathbf{d}}_2, \bar{\mathbf{d}}_3 \}^{-1}$. Leading to a general relation between frames for any vector $\mathbf{v}$: $\mathbf{v}=\mathbf{Q}\bar{\mathbf{v}}$, $\bar{\mathbf{v}}=\mathbf{Q}^T \mathbf{v}$, where $\bar{\mathbf{v}}$ denotes a vector in the lab frame and $\mathbf{v}$ is a vector in the local frame.  Here $s \in [0, L_0]$ is the material coordinate of a rod of rest-length $L_0$, $L$ denotes the deformed filament length and $t$ is time. If the rod is unsheared, $\bar{\mathbf{d}}_3$ points along the centerline tangent $\partial_s \bar{\mathbf{x}}=\bar{\mathbf{x}}_s$ while $\bar{\mathbf{d}}_1$ and $\bar{\mathbf{d}}_2$ span the normal--binormal plane. Shearing and extension shift $\bar{\mathbf{d}}_3$ away from $\bar{\mathbf{x}}_s$, which can be quantified with the shear vector $\boldsymbol{\sigma}=\mathbf{Q} (\bar{\mathbf{x}}_s-\bar{\mathbf{d}}_3) = \mathbf{Q}\bar{\mathbf{x}}_s - \mathbf{d}_3$ in the \emph{local} frame. The curvature vector $\boldsymbol{\kappa}$ encodes $\mathbf{Q}$'s rotation rate along the material coordinate $\partial_s \mathbf{d}_j = \boldsymbol{\kappa} \times \mathbf{d}_j$, while the angular velocity $\boldsymbol{\omega}$ is defined by $\partial_t \mathbf{d}_j = \boldsymbol{\omega} \times \mathbf{d}_j$. We also define the velocity of the centerline $\bar{\mathbf{v}} = \partial_t\bar{\mathbf{x}}$ and, in the rest configuration, the bending stiffness matrix $\mathbf{B}$, shearing stiffness matrix $\mathbf{S}$, second area moment of inertia $\mathbf{I}$, cross-sectional area $A$ and mass per unit length $\rho$. Then, the dynamics \cite{Gazzola:2018} of a soft slender body is described by: 

{\small
\begin{equation}
\rho A \cdot \partial_t^2 \bar{\mathbf{x}} = \partial_s \left( \frac{\mathbf{Q}^T \mathbf{S} \boldsymbol{\sigma}}{e} \right) + e\bar{\mathbf{f}}\label{eq:lin}
\end{equation}
\begin{equation}
\begin{aligned}
\frac{\rho \mathbf{I}}{e} \cdot \partial_t \boldsymbol{\omega} = \ & \partial_s \left( \frac{\mathbf{B} \boldsymbol{\kappa}}{e^3} \right) + \frac{\boldsymbol{\kappa} \times \mathbf{B} \boldsymbol{\kappa}}{e^3} + \left( \mathbf{Q}\frac{\bar{\mathbf{x}}_s}{e} \times \mathbf{S} \boldsymbol{\sigma} \right)\\
&+ \left( \rho \mathbf{I} \cdot \frac{\boldsymbol{\omega}}{e} \right) \times \boldsymbol{\omega} + \frac{\rho \mathbf{I} \boldsymbol{\omega}}{e^2} \cdot \partial_t e + e\mathbf{c}\label{eq:ang}
\end{aligned}
\end{equation}
\normalsize}%
where Eqs.~(\ref{eq:lin}, \ref{eq:ang}) represents linear and angular momentum balance at every cross section, $e = |\bar{\mathbf{x}}_s|$ is the local stretching factor, and $\bar{\mathbf{f}}$ and $\mathbf{c}$ are the external force and couple line densities, respectively.

The simulated snakes have length $L=0.35$ m, diameter $d=7.7$ mm, and uniform density $\rho = 1000$ kg/m\textsuperscript{3} to match measurements of milk snakes \cite{Hu:2009}. The Young's modulus of the filament representing the snake body is $E=1$ MPa \cite{Guo:2008} and the gravitational acceleration is $g=9.81$ m/s\textsuperscript{2}. Lateral muscular torques are applied to the filament through the term $\mathbf{c}$ in Eq. \ref{eq:ang}, and are determined so as to recover curvature profiles consistent with the planar snake model. The period of the lateral undulation is $2$ seconds and the forward friction ratio is $\mu_f=0.089$, resulting in $Fr=0.1$ for all simulations. Additional lifting muscular torques are enabled to produce the results of Fig.~\ref{Fig:Figure_Elastica}b-d, while non-body lifting snakes have only planar muscular activation. The simulation incorporates the same Coulomb friction model as in the planar snake model above. More details on our Cosserat rod model, discretization parameteres, and additional information regarding the different cases of Fig. \ref{Fig:Figure_Elastica} are available in the supplementary information.

\vspace{-15pt}
\section*{acknowledgement}
\vspace{-10pt}

\noindent
We thank Henry Astley for the useful discussions and careful proof-reading.

\bibliography{scibib.bib} 

\end{document}